\documentclass[aps,float,prd,psfig,amsmath,onecolumn,twocolumn]{revtex4-2}
\usepackage[dvipdfmx]{graphicx}
\usepackage{amsmath,amssymb,natbib,siunitx,booktabs,url}
\usepackage{color}
 \usepackage[normalem]{ulem}
\useunder{\uline}{\ul}{}

\newcommand{\beq}{\begin{equation}}
\newcommand{\beqa}{\begin{eqnarray}}
		  \newcommand{\eeq}{\end{equation}}
\newcommand{\eeqa}{\end{eqnarray}}

\begin{document}

\title{Higher Harmonics of Double White Dwarfs in the Centihertz Band: \\Linking LISA and DECIGO}

\author{Naoki Seto}
\affiliation{Department of Physics, Kyoto University, Kyoto 606-8502, Japan}

\date{\today}

\begin{abstract}
We investigate the detectability of post-Newtonian higher harmonics from Galactic double white dwarfs in the centihertz band ($\sim 0.01$ Hz). Using a synthetic population, we show that, unlike the quadrupole mode, higher harmonics remain undetectable with LISA except for rare nearby systems. In contrast, planned mid-band (decihertz) observatories such as DECIGO and BBO will be able to detect the third harmonic for about 10\% of inspiral binaries above $\sim 5$ mHz, enabling statistical constraints on mass ratios. These results highlight the successive roles of LISA and future
decihertz missions in establishing a coherent strategy for space-based
gravitational-wave astronomy.
\end{abstract}


\maketitle

\section{Introduction}
\label{sec:intro}
Short-period double white dwarfs (DWDs) are numerous sources of
low-frequency gravitational waves (GWs) in our Galaxy and represent one of the
primary targets for the Laser Interferometer Space Antenna (LISA)
\cite{Nelemans:2001nr,Ruiter:2007xx,nissanke2012gravitational,Lamberts:2019nyk,Korol:2021pun,amaro2023astrophysics,toubiana2024interacting}.
Most DWDs are expected to circularize during their evolutionary history,
and their quadrupole emission (the fundamental $k=2$ harmonic) will be detected
from thousands of Galactic systems, providing a nearly complete census of the
population above $\sim 4$ mHz \cite{amaro2023astrophysics}.
While the quadrupole mode ensures detections across the Galaxy, 
it alone does not provide direct information on the binary mass ratio.

Even for circular binaries, higher harmonics arise at post-Newtonian (PN) order
and offer a direct probe of mass asymmetry
\cite{kidder1995coalescing,blanchet2014gravitational,poisson2014gravity}.
In particular, the first ($k=1$) and third ($k=3$) harmonics appear at
0.5PN order with amplitudes proportional to $\beta\Delta$,
where $\beta \sim v/c$ is the PN velocity parameter and $\Delta$ is the
fractional mass difference. These harmonics vanish  in the equal-mass
limit and therefore carry unique information about the distribution of mass
ratios in the binaries. Their importance has already been demonstrated
for black hole binaries observed with ground-based interferometers, where they
significantly improve parameter estimation
\cite{abbott2020gw190412,abbott2020gw190814}.
For Galactic DWDs, however, the orbital velocities are small ($\beta < 10^{-2}$),
and the amplitudes are suppressed by a factor $\sim\beta$ relative to the
quadrupole mode.

{From a theoretical perspective,
for DWD systems, the mass ratio is the critical parameter for  the stability of
Roche--lobe overflow and thus closely related to whether a system undergoes a
merger or evolves into a long--lived AM~CVn binary
\citep[e.g.][]{PostnovYungelson2014}. At the population level, the mass asymmetry distribution provides a key
diagnostic of DWD formation channels  \citep{amaro2023astrophysics}.
The detection of higher harmonics enables a direct measurement of the mass
asymmetry and thereby provides observational access to these aspects.
}

Recently, Ref.~\cite{Seto:2025vfg} presented the first explicit study of
PN harmonics from DWDs and showed that they might become observable with LISA
in favorable cases. By hypothetically placing two representative systems (HM Cnc and
ZTF J1539 \cite{chakraborty2024expanding}) at different distances, that study
examined the prospects for detection with LISA and with the combined
LISA--Taiji--TianQin (LTT) network \cite{hu2017taiji,luo2016tianqin,Cai:2023ywp}.
It concluded that while most Galactic DWDs are too weak in their higher harmonics,
there remains a non-negligible chance that nearby systems could yield a detection,
suggesting that LISA might provide the first glimpse of PN harmonics from DWDs.

The present paper extends Ref.~\cite{Seto:2025vfg} within a systematic and statistical
framework. Rather than focusing on a few known binaries, we construct a
synthetic Galactic population by sampling masses, frequencies,
inclinations, and spatial locations. This enables us to evaluate the
detectability of PN harmonics in a more realistic setting. We further
extend the analysis to future decihertz observatories such as DECIGO
\cite{Kawamura:2011zz} and BBO \cite{Harry:2006fi}, whose improved
sensitivity in the $\sim 0.01$--0.1 Hz range allows them to detect
higher harmonics from hundreds of systems. 
While LISA will provide the
quadrupole detections necessary for a Galactic census, decihertz
observatories will capitalize on this foundation by extracting
mass-ratio information from the higher harmonics. Together, these
missions establish a sequential observational path for space-based
gravitational-wave astronomy, extending coverage from the millihertz to
the decihertz regime.

The remainder of this paper is organized as follows. In Sec.~\ref{sec:signal}
we introduce the waveform model including the higher harmonics appearing
at 0.5PN order. Sec.~\ref{sec:population} presents the construction of the
synthetic population. In Sec.~\ref{sec:obs_impact} we evaluate the
detectability of the harmonics with LISA and DECIGO. Sec.~\ref{sec:discussion}
discusses the astrophysical implications and synergies with
electromagnetic observations. Finally, Sec.~\ref{sec:conclusions} provides
our main conclusions.

\section{Signal model}
\label{sec:signal}

\subsection{Waveform}

We consider quasi-circular DWDs with component masses $m_a,m_b$ ($m_a>m_b$),
total mass $M=m_a+m_b$, symmetric mass ratio $\eta=m_am_b/M^2$, and chirp
mass $\mathcal{M}=M\,\eta^{3/5}$. We also define the fractional mass as
difference
\begin{equation}
 \Delta=\frac{m_a-m_b}{M}.
\end{equation}
The orbital frequency is $f_{\rm orb}$, and the gravitational-wave (GW)
harmonics are $f_k = k f_{\rm orb}$, with the fundamental ($k=2$)
harmonic at $f_2=2f_{\rm orb}$. In this paper we refer to $k=1,2,3$ as
the \emph{first}, \emph{fundamental}, and \emph{third} harmonics,
respectively; when referring to $\{k=1,3\}$ collectively, we use the term
\emph{odd harmonics}.

We introduce the post-Newtonian (PN) velocity parameter, $\beta\sim v/c$,
defined by
\begin{align}
 \beta &\equiv \left(\frac{\pi G M f_2}{c^3}\right)^{1/3}, \label{eq:beta}\\
 &\simeq 5.37\times10^{-3}
   \left(\frac{M}{M_\odot}\right)^{1/3}
   \left(\frac{f_2}{10\,\mathrm{mHz}}\right)^{1/3}. \nonumber
\end{align}

At 0.5PN order, the two polarizations in the principal polarization
frame take the form
\cite{kidder1995coalescing,blanchet2014gravitational,poisson2014gravity} as
\begin{align}
 h_{+}(t) &= A \Big[ a_{2,+}(I)\cos(2\Psi)  \nonumber \\
 &\quad + \beta \Delta \big( a_{1,+}(I)\cos\Psi
     + a_{3,+}(I)\cos(3\Psi) \big) \Big], \label{eq:hplus} \\
 h_{\times}(t) &= A \Big[ a_{2,\times}(I)\sin(2\Psi) \nonumber \\
 &\quad + \beta \Delta \big( a_{1,\times}(I)\sin\Psi
     + a_{3,\times}(I)\sin(3\Psi) \big) \Big]. \label{eq:hcross}
\end{align}
where
\begin{equation}
 A = \frac{2 (G\mathcal{M})^{5/3} (\pi f_2)^{2/3}}{c^4 d}
 \label{eq:quad_amp}
\end{equation}
is the leading-order quadrupole amplitude at distance $d$, $\Psi$ is the
orbital phase with $\dot{\Psi}=2\pi f_{\rm orb}$, and $I$ is the
inclination.

The angular coefficients $a_{k,\{+,\times\}}(I)$ are
\begin{align}
a_{2,+}(I) &= -\bigl(1+\cos^2 I\bigr), &
a_{2,\times}(I) &= -2\cos I, \\
a_{1,+}(I) &= -\frac{\sin I}{8}\bigl(5+\cos^2 I\bigr), &
a_{1,\times}(I) &= -\frac{3}{4}\sin I\cos I, \\
a_{3,+}(I) &= +\frac{9}{8}\sin I\bigl(1+\cos^2 I\bigr), &
a_{3,\times}(I) &= +\frac{9}{4}\sin I\cos I .
\end{align}
In Eqs.~(\ref{eq:hplus}) and (\ref{eq:hcross}), the fundamental mode
with phase $2\Psi$ provides the dominant signal, while the first and third
harmonics with phases $\Psi$ and $3\Psi$ appear only at 0.5PN order and
scale as $\propto \beta\Delta$. They vanish in the equal-mass limit
$\Delta=0$ and therefore provide a direct probe of mass asymmetry.

\begin{figure}[t]
  \centering
  \includegraphics[width=0.9\linewidth]{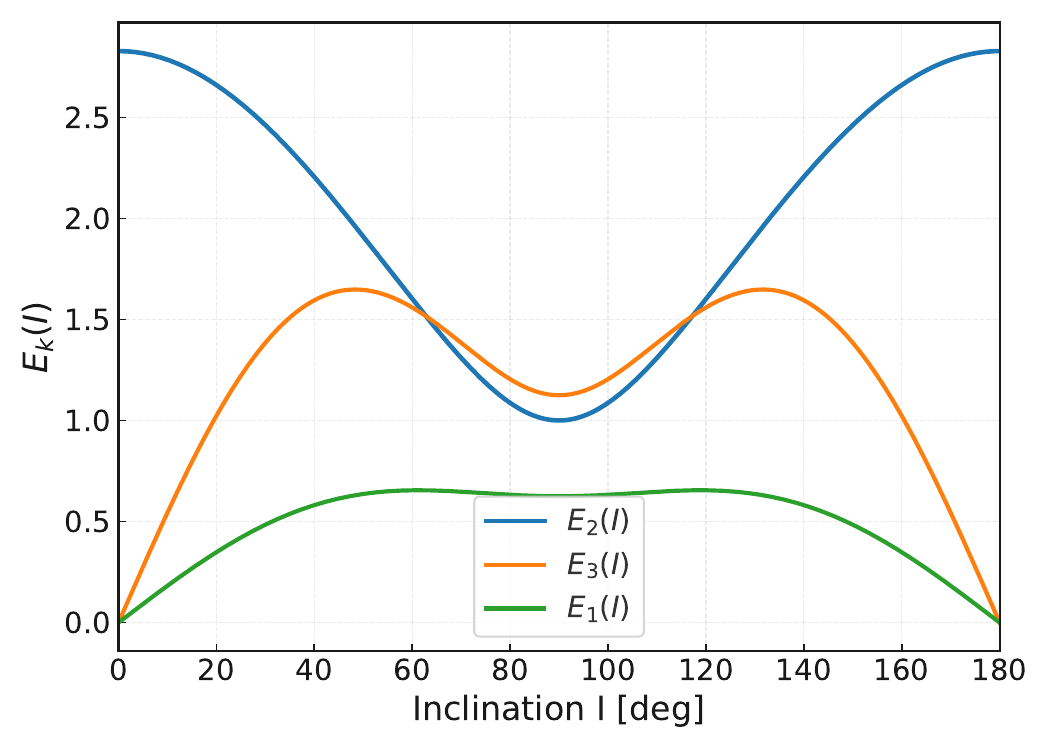}
  \caption{Inclination dependence of the geometrical factors
  $E_k(I)=\sqrt{a_{k,+}^2+a_{k,\times}^2}$ for $k=1,2,3$. The odd-harmonic
  factors vanish for face-on configurations ($I=0$ or $\pi$).}
  \label{fig:EkI}
\end{figure}

For convenience, we define the geometrical factor
\begin{equation}
E_k(I) \equiv \sqrt{a_{k,+}^2(I)+a_{k,\times}^2(I)},
\label{eq:Ek}
\end{equation}
which characterizes the effective angular strength of each harmonic $k$.
Figure~\ref{fig:EkI} shows $E_k(I)$ for $k=1,2,3$; the odd-harmonic factors
vanish for face-on configurations ($I=0,\pi$). We also have  
\begin{equation}
\frac{E_3(I)}{E_2(I)}=\frac{9}{8}\sin I,
\end{equation}
and
\begin{equation}
\frac13\le \frac{E_1(I)}{E_3(I)}\le \frac59. \label{ineq}
\end{equation}

The overall scaling of the odd-harmonic amplitude is $A\,\beta\,\Delta$.
At fixed $M$, we write $m_a=M(1+\Delta)/2$ and $m_b=M(1-\Delta)/2$,
which gives $\eta=(1-\Delta^2)/4$. Since $A\propto \mathcal{M}^{5/3}
\propto \eta$ (for fixed $M,f_2,d$) and $\beta\propto M^{1/3}$ is
independent of $\Delta$, we obtain
\begin{equation}
A\,\beta\,\Delta \;\propto\; \frac{\Delta\,(1-\Delta^2)}{4},
\end{equation}
which is maximized at $\Delta=1/\sqrt{3}\simeq 0.58$, corresponding to a
moderately asymmetric binary ($m_a:m_b\simeq 3:1$).

In addition to the first and third harmonics arising at 0.5PN order,
further contributions exist from eccentricity-induced harmonics and from
the 1PN corrections. The former have already been discussed in detail
elsewhere \cite{Seto:2025vfg} and are not repeated in this paper.
The latter scale as $\beta^2$ and are therefore expected to be far weaker
than the 0.5PN harmonics considered here. A quantitative assessment of
their impact is deferred to Sec.~\ref{sec:obs_impact}.

\subsection{Signal-to-noise ratios}

The response of a detector to the incoming waveforms
(\ref{eq:hplus}) and (\ref{eq:hcross}) also depends on the
sky position and polarization angle. For LISA-like orbital
configurations, however, it is well approximated by the
sky- and polarization-averaged response \cite{yagi2011detector}.
For an integration time that is an integer multiple of one year,
the minimum response to an arbitrary linear polarization is
$0.89$ times the root-mean-square value
\cite{Seto:2004ji,yagi2011detector}. In the following we adopt
this averaged treatment, retaining only the inclination
dependence through $E_k(I)$ \footnote{This approximation is not valid
for the proposed TianQin orbit.}.

For a quasi-monochromatic source integrated over time $T$, the
per-harmonic SNRs in a single LISA-like unit are given by
\cite{Robson:2018ifk}
\begin{align}
 \rho_2 &\simeq \frac{A\,E_2(I)}{\sqrt{S_n(f_2)}}\sqrt{T}, \label{eq:rho2}\\[4pt]
 \rho_3 &\simeq \frac{A\,\beta\Delta\,E_3(I)}{\sqrt{S_n(f_3)}}\sqrt{T}, \label{eq:rho3}\\[4pt]
 \rho_1 &\simeq \frac{A\,\beta\Delta\,E_1(I)}{\sqrt{S_n(f_1)}}\sqrt{T}. \label{eq:rho1}
\end{align}
Here $S_n(f)=S_{\rm inst}(f)+S_{\rm exgal}(f)$ is the
angle-averaged noise power spectral density, consisting of the
instrumental part $S_{\rm inst}(f)$ and the extra-galactic
white-dwarf confusion foreground $S_{\rm exgal}(f)$.
{Galactic confusion noise from unresolved binaries is not included,
since our analysis is restricted to binaries with
$f_2 \gtrsim 5\,\mathrm{mHz}$, where such systems are expected
to be individually resolved by LISA.
}

For $S_{\rm inst}(f)$ we adopt the analytic form of
\cite{Robson:2018ifk} for LISA, those of
\cite{2020ResPh..1602918L,TianQin:2020hid} for Taiji and TianQin,
and the formulas in \cite{yagi2011detector} for DECIGO and BBO.
Based on \cite{Farmer:2003pa}, we fit the extra-galactic DWD
foreground around $1$--$60\,\mathrm{mHz}$ by
\begin{equation} \label{exg}
  S_{\rm exgal}(f) = \left[\,0.5 \times 10^{\,a + b u + c u^2}\,\right]^2,
\end{equation}
with $u = \log_{10}(f/10^{-2}\,{\rm Hz})$ and
$(a,b,c)=(-20.523,\,-1.847,\,-0.722)$. The prefactor $0.5$
corresponds to the pessimistic scaling discussed in
\cite{Farmer:2003pa}.

\begin{figure*}[t]
  \centering
  \includegraphics[width=0.65\textwidth]{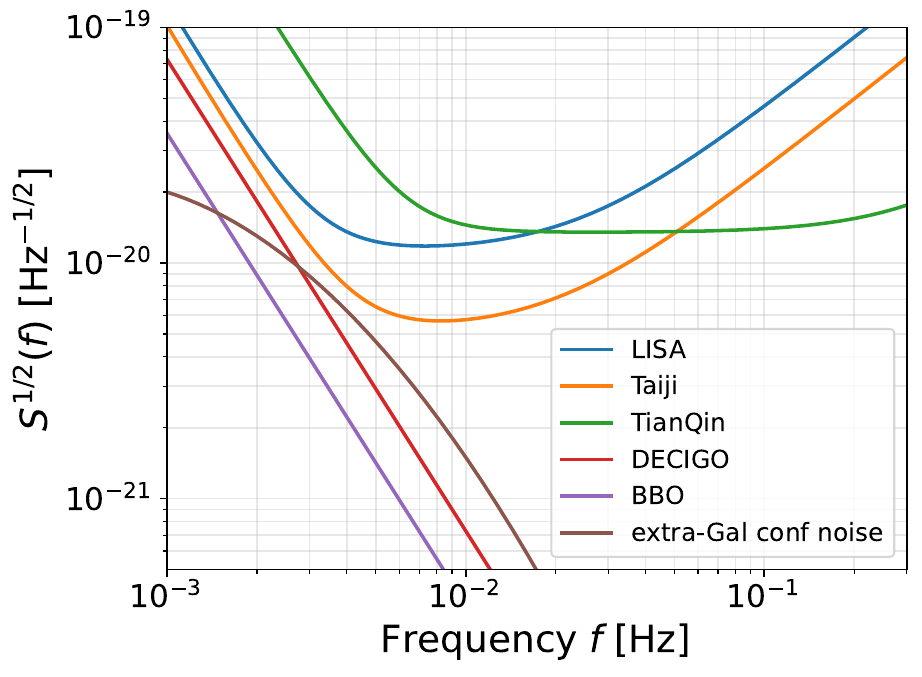}
  \caption{Adopted noise spectra for space-based detectors.
  The curves show the instrumental terms $S_{\rm inst}(f)$
  (LISA, Taiji, TianQin, DECIGO, BBO) and the extra-galactic
  white-dwarf foreground $S_{\rm exgal}(f)$ from Eq.~(\ref{exg}).
  The vertical axis shows $\sqrt{S(f)}$ (Hz$^{-1/2}$), where $S(f)$ denotes the respective spectra.
  The figure displays $S_{\rm inst}$ and $S_{\rm exgal}$ separately;
  it does not show the summed $S_n=S_{\rm inst}+S_{\rm exgal}$ nor the
  $S_n/3$ scaling used for the three independent DECIGO/BBO units.}
  \label{fig:noise_exgal}
\end{figure*}

Figure~\ref{fig:noise_exgal} illustrates the adopted noise spectra.
Around $10\,\mathrm{mHz}$, LISA, Taiji, and TianQin show relatively weak
frequency dependence in their sensitivities, dominated by position
noise ($\propto f^0$). In contrast, for DECIGO and BBO the sensitivity
is dominated by the extra-galactic foreground, producing a much sharper
frequency dependence. This distinction is crucial for evaluating the
detectability of higher harmonics at $f\gtrsim 10\,\mathrm{mHz}$.

The nominal designs of DECIGO and BBO include four units
\cite{Kawamura:2011zz}. Two of them are co-located at the same site
and therefore do not provide independent noise realizations.
The remaining three units are separated by light-travel times of
order $10^3\,\mathrm{s}$, sufficient to regard their confusion noises
as uncorrelated around $10\,\mathrm{mHz}$ \cite{Seto:2021crt}.
Throughout this work we therefore take the effective number of
independent detectors to be three, and consistently use the
effective sensitivity $S_n(f)/3$.

{In this paper, 
we adopt $\rho_k=5$ as a conservative working threshold for higher
harmonics ($k=1,3$). In contrast to blind searches, which typically require
higher thresholds (e.g., close to $7$ \cite{DigmanCornish2022}), the present search
is a targeted analysis, once the dominant quadrupole mode is identified
and most of the source parameters are tightly constrained.
}

\section{Population setup}
\label{sec:population}
In this section we describe the construction of the DWD sample used in our
analysis, following the scheme of \cite{Seto:2022iuf}. 
{
Because the total number of Galactic DWDs in the relevant
frequency range is subject to astrophysical uncertainties, our analysis
primarily emphasizes fractions of systems rather than absolute source counts.
To this end, we construct a representative Monte Carlo population of
4000  binaries, which
serves as a parent sample for the subsequent analysis.
}

More specifically, we generate a steady-state population by injecting inspiral DWDs at
$f_{2,\min}=4$ mHz, corresponding to the flux of binaries that enter the
band, and then evolve them under gravitational radiation reaction. Our focus
is restricted to the inspiral phase. Detached systems lose orbital
angular momentum mainly through gravitational radiation and gradually
inspiral until the less massive component fills its Roche lobe. At that point mass transfer begins, and the subsequent evolution depends
on the stability of the transfer \cite{1967AcA....17..287P}:
if unstable the binary merges, whereas if stable it enters an outspiraling
phase \cite{amaro2023astrophysics}. Since the physics of mass-transfer
stability remains uncertain (see, e.g., \cite{Marsh:2003rd,Gokhale:2006yn}),
we restrict our attention to the pre-contact inspiral regime, which is well
described by gravitational radiation reaction. This
restriction is further justified by the fact that GW amplitude
measurements, unlike phase evolution, are not directly linked to the
sign of $\dot f$.

In Sec.~\ref{sec:mass_dist} we discuss the distribution of component
masses. In Sec.~\ref{sec:freq_dist} we specify the assignment of orbital
frequencies, and in Sec.~\ref{sec:galaxy_dist} we introduce the Galactic
spatial distribution of DWDs. This framework provides the baseline
sample from which signal-to-noise ratios and detection statistics are
evaluated in the following sections.

\subsection{Binary injection} \label{sec:mass_dist}

We inject inspiraling DWDs at $f_{2,\min}=4$ mHz, representing the
inflow of systems into the observational band, and then draw their
component masses $m_a$ and $m_b$ independently from the same
one-body distribution $P(m)$. We order them such that $m_b<m_a$.
This choice implicitly determines the distribution of mass ratios
without introducing it as an explicit parameter.

We approximate the one-body distribution $P(m)$ by a two-component
Gaussian mixture with peaks at $0.24\,M_\odot$ and $0.64\,M_\odot$
(representing typical He and CO white dwarfs), with relative fractions
of 0.55 and 0.45, respectively (see, e.g., \cite{2025MNRAS.541.3494M}
for a recent observational study). The support is restricted to
$0.11$--$1.1\,M_\odot$, and $P(m)$ is normalized to unity over this
interval. 

The instantaneous (snapshot) mass distribution is biased toward
long-lived systems. In the inspiral phase the quadrupole frequency
evolves as
\begin{equation}
  \dot f_2 \propto \mathcal{M}^{5/3} f_2^{11/3},
  \label{eq:fdot}
\end{equation}
where $\mathcal{M}$ is the chirp mass (ignoring the finite size effects). The corresponding residence time
in a frequency band scales approximately as $\mathcal{M}^{-5/3}$.
Thus, our snapshot-level mass distribution is given by 
\begin{equation}
  p_{\rm snap}(m_a,m_b) \propto
  \mathcal{M}^{-5/3}\,P(m_a)\,P(m_b).
  \label{eq:snap_mass}
\end{equation}

\subsection{Frequency assignment}
\label{sec:freq_dist}

We set the upper edge of the inspiral band at the onset of mass transfer
from the donor, when its radius becomes equal to the Roche-lobe radius
$R_L$. For simplicity we adopt the approximate scaling of
\cite{1967AcA....17..287P},
\begin{equation}
  R_L \simeq 2\,3^{-4/3}\,a\,m_b^{1/3}(m_a+m_b)^{-1/3},
  \label{eq:RL}
\end{equation}
where $a$ is the orbital separation. For the donor radius we use the
analytic expression for a completely degenerate white dwarf given in
Verbunt \& Rappaport \cite{1988ApJ...332..193V}, multiplied by an
inflation factor $\eta_R=1.2$ to allow for possible radius expansion
\footnote{This choice is consistent with the Roche-lobe filling of
HM~Cnc for $m_b=0.17\,M_\odot$ at $f_2=6.2$ mHz \cite{munday2023two}.}

The corresponding binary separation $a_{\rm RL}(m_a,m_b)$ then yields
the limiting quadrupole frequency
\begin{equation}
  f_{2,{\rm RL}}(m_a,m_b)
  = \frac{1}{\pi}\left[\frac{G(m_a+m_b)}{a_{\rm RL}^3}\right]^{1/2}.
  \label{eq:fRL}
\end{equation}

{Given a binary with component masses $(m_a,m_b)$, we assign its GW
frequency by assuming a steady--state inspiral driven by radiation
reaction, for which $dN/df_2 \propto f_2^{-11/3}$.
We therefore draw $f_2$ from the interval
\begin{equation}
  f_2 \in \bigl[f_{2,\min},\,f_{2,{\rm RL}}(m_a,m_b)\bigr],
  \label{eq:freq_range}
\end{equation}
where the lower cutoff is fixed at the injection frequency
$f_{2,\min}=4~\mathrm{mHz}$.
The normalized probability density is then
\begin{equation}
  p_{\rm inj}(f_2|m_a,m_b)
  = \frac{8}{3}\,
    \frac{f_2^{-11/3}}
         {f_{2,\min}^{-8/3}-f_{2,{\rm RL}}(m_a,m_b)^{-8/3}}.
  \label{eq:pf_inj}
\end{equation}}

{In the subsequent analysis, we restrict the working sample to
$f_2 \gtrsim 5\,\mathrm{mHz}$ to focus on a clean regime where Galactic
binaries are safely resolvable and less affected by our artificial injection freqeuncy of 4mHz.  As demonstrated below,  the qualitative conclusions are not
sensitive to modest variations of this threshold frequency.}

\begin{figure}[t]
  \centering
  \includegraphics[width=0.47\textwidth]{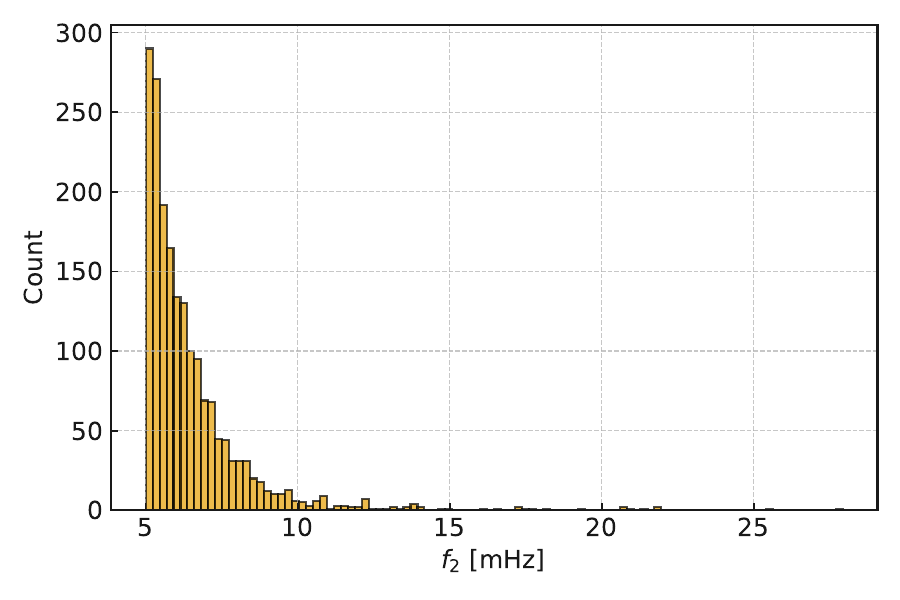}
  \caption{Histogram of quadrupole GW frequencies $f_2$ for the
  1858 binaries with $f_2 \geq 5$ mHz in the analysis sample.}
  \label{fig:freq_dist}
\end{figure}

Out of 4000 binaries injected at $f_{2,\min}=4$mHz, a total of 1858 systems evolve to
$f_2 \geq 5$ mHz. These constitute the working sample used throughout this paper. 
The corresponding frequency distribution is shown in Fig.~\ref{fig:freq_dist}. 
The sample contains 73 systems at $f_2 \geq 10$ mHz and 16 systems at $f_2 \geq 15$ mHz.

\begin{figure}[t]
  \centering
  \includegraphics[width=0.47\textwidth]{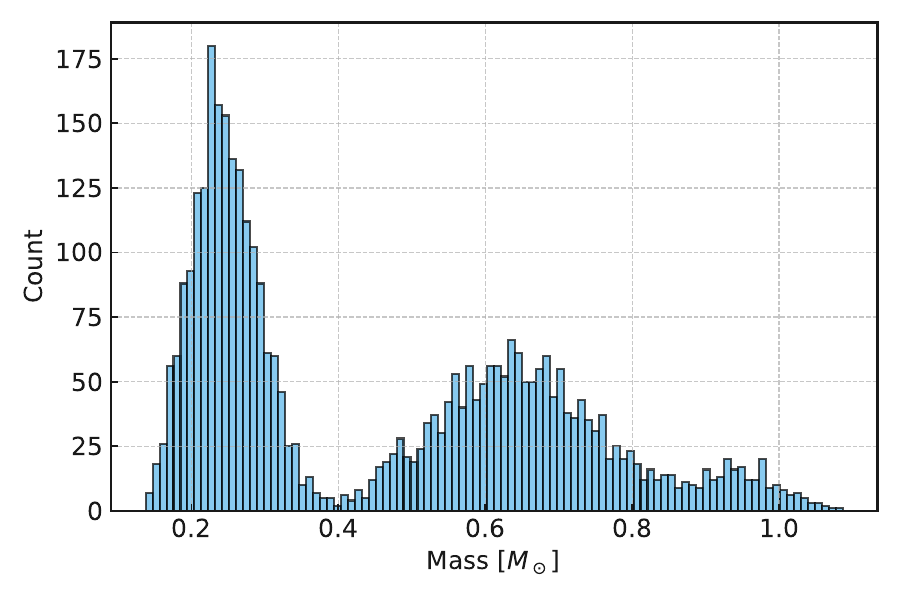}
  \caption{Histogram of component masses for the  1858 binaries.
  Distinct peaks correspond to He and CO white dwarfs, with a high-mass
  tail associated with the CO component.}
  \label{fig:mass_dist}
\end{figure}

In addition, Fig.~\ref{fig:mass_dist} shows the component-mass
distribution for the same set of 1858 DWDs (3716 white dwarfs in total).
The He and CO peaks are clearly visible, together with a small high-mass
tail associated with the CO component.

\subsection{Galactic distribution}
\label{sec:galaxy_dist}

We next describe the spatial distribution of Galactic close DWDs, modeled
as an axisymmetric thin disk plus a spherical bulge component
\cite{nissanke2012gravitational}. We place the observer at
$(R_0,\phi,z)=(8.3~\mathrm{kpc},0,0)$ in Galactocentric cylindrical
coordinates.

The normalized source density is written as
\begin{equation}
  n(R,z) = (1-f_{\rm bulge})\,n_{\rm disk}(R,z)
         + f_{\rm bulge}\,n_{\rm bulge}(r),
  \label{eq:gal_density}
\end{equation}
where $n_{\rm disk}$ and $n_{\rm bulge}$ are each normalized to unity,
$f_{\rm bulge}$ denotes the adopted bulge fraction, and
$r=\sqrt{R^2+z^2}$. We set $f_{\rm bulge}=1/3$.

The disk profile is modeled by a double exponential,
\begin{equation}
  n_{\rm disk}(R,z) =
   \frac{1}{4\pi R_d^2 z_d}\,
   \exp\!\left(-\frac{R}{R_d}\right)\,
   \exp\!\left(-\frac{|z|}{z_d}\right),
  \label{eq:disk_profile}
\end{equation}
and the bulge by a spherical exponential,
\begin{equation}
  n_{\rm bulge}(r) =
   \frac{1}{8\pi R_b^3}\,
   \exp\!\left(-\frac{r}{R_b}\right).
  \label{eq:bulge_profile}
\end{equation}
We adopt $R_d=2.5~\mathrm{kpc}$ for the disk radial scale length,
$z_d=0.2~\mathrm{kpc}$ for the vertical scale height, and
$R_b=1.0~\mathrm{kpc}$ for the bulge scale length.

The heliocentric distance to a point $(R,\phi,z)$ is
\begin{equation}
  d(R,\phi,z) =
  \sqrt{R^2+R_0^2 - 2RR_0\cos\phi + z^2},
  \label{eq:distance}
\end{equation}
and the cumulative distribution of source distances is
\begin{equation}
  F(<d) = \int_{d(R,\phi,z)\le d} n(R,z)\,R\,dR\,d\phi\,dz .
  \label{eq:cdf_def}
\end{equation}

\begin{figure}[t]
  \centering
  \includegraphics[width=0.9\linewidth]{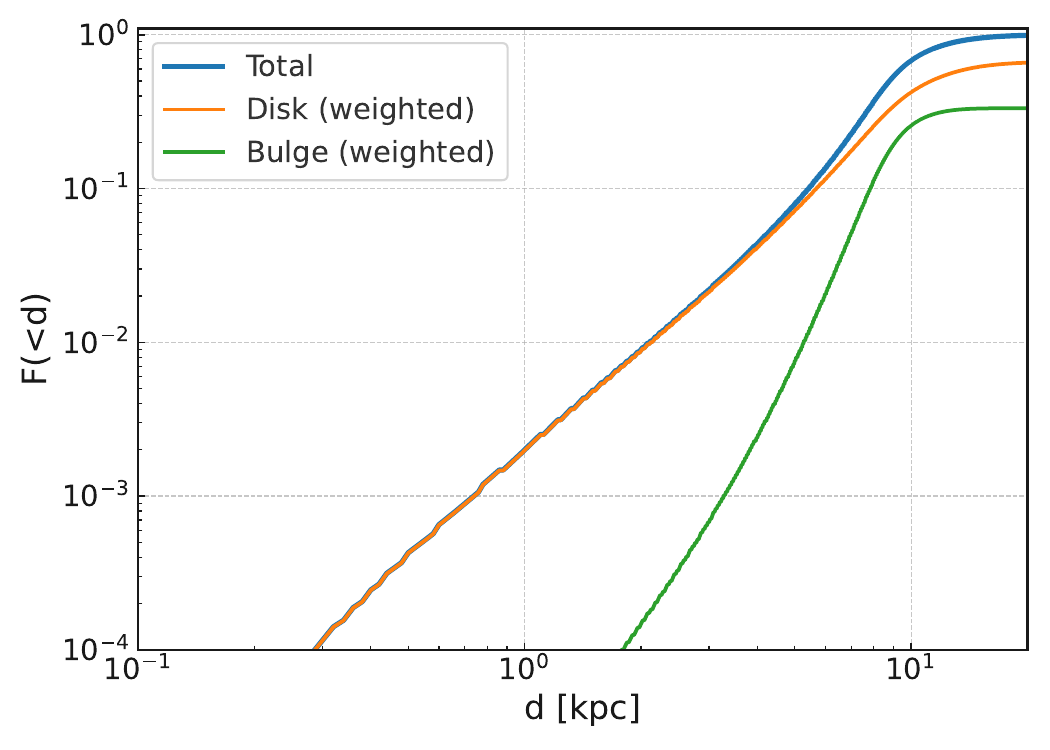}
  \caption{Cumulative distribution of heliocentric distances $F(<d)$
  for the disk, bulge, and total populations defined in
  Eq.~(\ref{eq:cdf_def}).}
  \label{fig:distance_cdf}
\end{figure}

Figure~\ref{fig:distance_cdf} shows the cumulative distribution. 
At small distances ($d\lesssim0.5$ kpc), the cumulative distribution 
approximately follows $F(<d)\propto d^{2.5}$, reflecting the local geometry 
of the thin disk \cite{Seto:2025vfg}.
 The nearby population is dominated by the
disk, while the bulge contribution becomes significant only toward the
Galactic center. At larger radii the disk again dominates, and nearly
all Galactic DWDs are contained within $d<20$ kpc in this model.

The binary inclination $I$ is assumed to be isotropic, with $\cos I$
uniformly distributed in $[-1,1]$.

\section{Observational Impact}
\label{sec:obs_impact}

In this section we quantify the detection prospects for different
harmonics of the GW signal from Galactic DWDs. Our aim is to clarify
the successive stages of observation, beginning with LISA-class detectors
and extending to future decihertz observatories. We first present overall
detection counts, then examine the quadrupole baseline, turn to the odd
harmonics, and finally study their dependence on distance and frequency
before closing with a brief summary.

\subsection{Overview}
\label{sec:overview}

LISA (and the LISA--Taiji--TianQin (LTT) network) can detect the
fundamental quadrupole mode with high confidence, providing an initial
inventory of nearly monochromatic Galactic binaries. DECIGO and BBO,
with superior sensitivity in the 0.01--0.1 Hz band, can probe the
higher harmonics. Since their effective sensitivities are limited by the
extra-galactic confusion foreground, DECIGO and BBO yield similar
performance (see Fig.~\ref{fig:noise_exgal}), and we mainly present
results for DECIGO in what follows. The potential role of the LTT
network in enabling the first detection of the third harmonic was previously noted
by \cite{Seto:2025vfg}, but its impact on the systematic study
presented here is very limited.

{
Table~\ref{tab:snr_counts} summarizes the number of binaries with
$\rho_k>5$ (the SNR of the $k$th harmonic) across representative
detectors.}

\begin{table}[t]
  \centering
  \caption{Number of binaries with ${\rho_k}>5$ for each harmonic
  ($k=1,2,3$) and detector configuration. The analysis sample contains
  1858 systems at $f_2 \ge 5$~mHz. LTT denotes the LISA--Taiji--TianQin
  network.}
  \vspace{0.5em}
\begin{tabular}{lcccc}
    \hline\hline
    Harmonic & LISA (4yr) & LTT (4yr) & DECIGO (10yr) & BBO (10yr) \\
    \hline
    $\rho_{1}$ & 0 & 0 & 1 & 1 \\
    $\rho_{2}$ & 1857 & 1858 & 1858 & 1858 \\
    $\rho_{3}$ & 0 & 1 & 135 & 170 \\
    \hline\hline
  \end{tabular}
  \label{tab:snr_counts}
\end{table}
\subsection{Quadrupole performance}
\label{sec:rho2_performance}

\begin{figure}[t]
  \centering
  \includegraphics[width=0.47\textwidth]{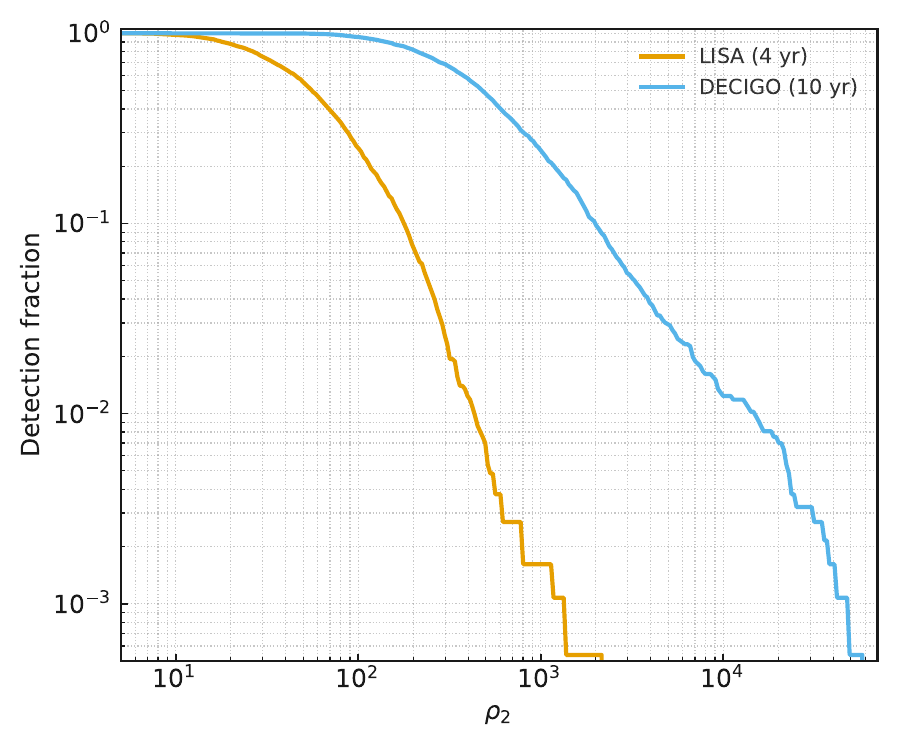}
  \caption{Detection fraction of the quadrupole harmonic ($\rho_2$) as a
  function of the SNR threshold for LISA (orange) and DECIGO (blue).}
  \label{fig:detfrac_rho2}
\end{figure}

Figure~\ref{fig:detfrac_rho2} shows the detection fraction of the
quadrupole harmonic ($\rho_2$) as a function of the SNR threshold.
For the LISA mission with a 4 yr observation time, nearly all of the
1858 binaries lie above the threshold. The minimum quadrupole SNR is $\rho_2 \simeq 4.7$.
{Only one system has $\rho_2 \le 5$, and only three systems have
$\rho_2 \le 7$ among the 1858 binaries considered.
Thus, even adopting a higher threshold of $\rho_2 = 7$  for
blind searches, LISA would still robustly detect the fundamental modes
for essentially the entire Galactic DWD population in this frequency
range.}

 The distribution extends up to
$\rho_2 \simeq 2200$, demonstrating a wide dynamic range. This demonstrates
that LISA can robustly detect virtually all Galactic DWDs in this
frequency range, consistent with the long-standing conclusion that LISA
will provide a nearly complete census of the population for
$f_2\gtrsim 4$ mHz \cite{amaro2023astrophysics}.

For DECIGO, operating with a 10 yr mission, the sensitivity is even
higher. The weakest source in our sample still has $\rho_2 \simeq 31$,
while the strongest reaches $\rho_2 \simeq 5.9\times 10^{4}$. Thus the
detection is guaranteed with ample margin.

\begin{figure*}[t]
  \centering
  \includegraphics[width=0.48\textwidth]{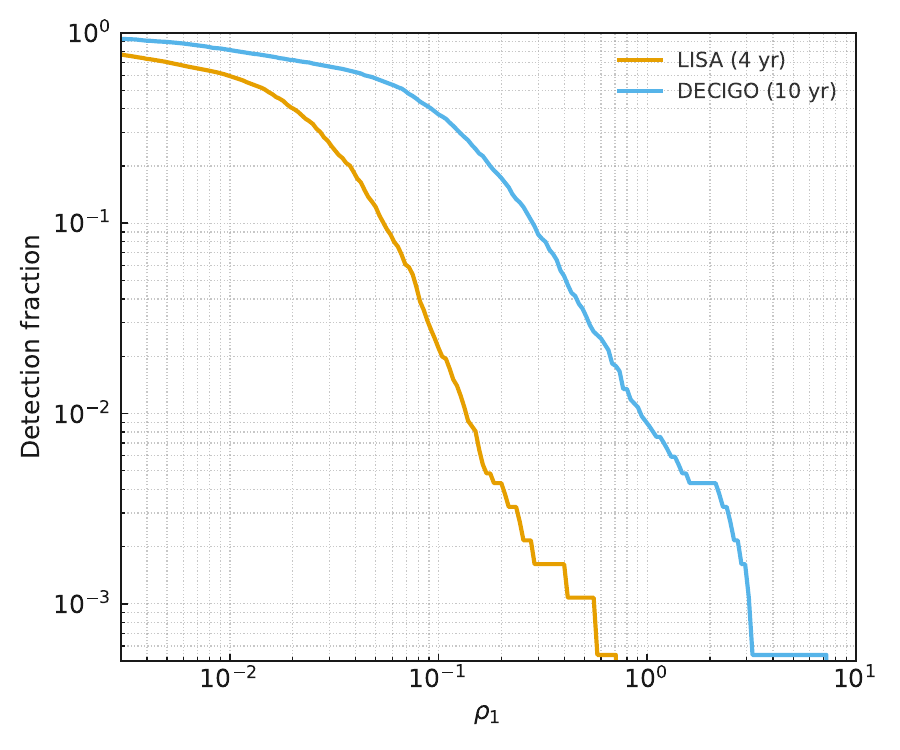}
  \includegraphics[width=0.48\textwidth]{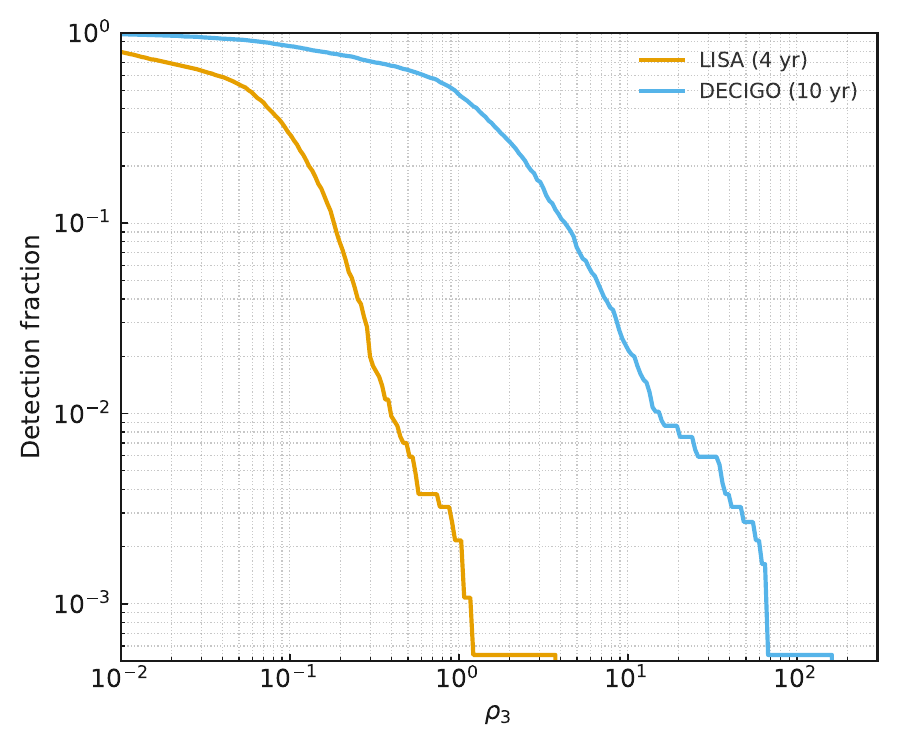}
  \caption{Detection fraction as a function of the SNR threshold for the
  first ($\rho_1$, left) and third ($\rho_3$, right) harmonics, for LISA
  (orange) and DECIGO (blue).}
  \label{fig:detfrac_rho1_rho3}
\end{figure*}

\subsection{Odd harmonics}
\label{sec:odd_harmonics}

Figure~\ref{fig:detfrac_rho1_rho3} shows the detection fractions for the
first ($\rho_1$) and third ($\rho_3$) harmonics as a function of the SNR
threshold. For LISA, the odd harmonics are effectively inaccessible: no
systems exceed ${\rm SNR}=5$ for either $\rho_1$ or $\rho_3$, and only a
single case appears in the LTT configuration (4 yr). As discussed in
\cite{Seto:2025vfg}, rare nearby binaries might yield detectable odd
harmonics even with LISA, but such cases would be exceptional rather
than representative. LISA is therefore unlikely to provide a systematic
view of the odd harmonics.

{
For DECIGO, we explicitly examine how the detectability of the third
harmonic depends on the adopted SNR threshold $\rho_3$.
Figure~7 (right panel) shows the detection fraction as a function of
this threshold.
Lowering the threshold from the fiducial value $\rho_3=5$ to $4$ and
$3$ increases the number of detectable systems from 135 to 197 and 306,
respectively, in our
sample of 1858 binaries.
Although the absolute detection counts vary, the qualitative trends remain
unchanged.
}

For completeness, we also consider the combined odd--harmonic SNR
$\rho_{\rm odd} \equiv \sqrt{\rho_1^2+\rho_3^2}$. Among the 1858 sources,
the number of systems with $\rho_{\rm odd}>5$ is 0 (LISA, 4 yr),
1 (LTT, 4 yr), 137 (DECIGO, 10 yr), and 172 (BBO, 10 yr). Compared with
the corresponding counts for the third harmonic alone (0, 1, 135, and
170), the contribution of the first harmonic remains marginal, though it
can slightly increase the detection statistics for the decihertz
observatories.

\subsection{Spatial and frequency domains}
\label{sec:rho3_scatter}

\begin{figure*}[t]
  \centering
  \includegraphics[width=0.48\textwidth]{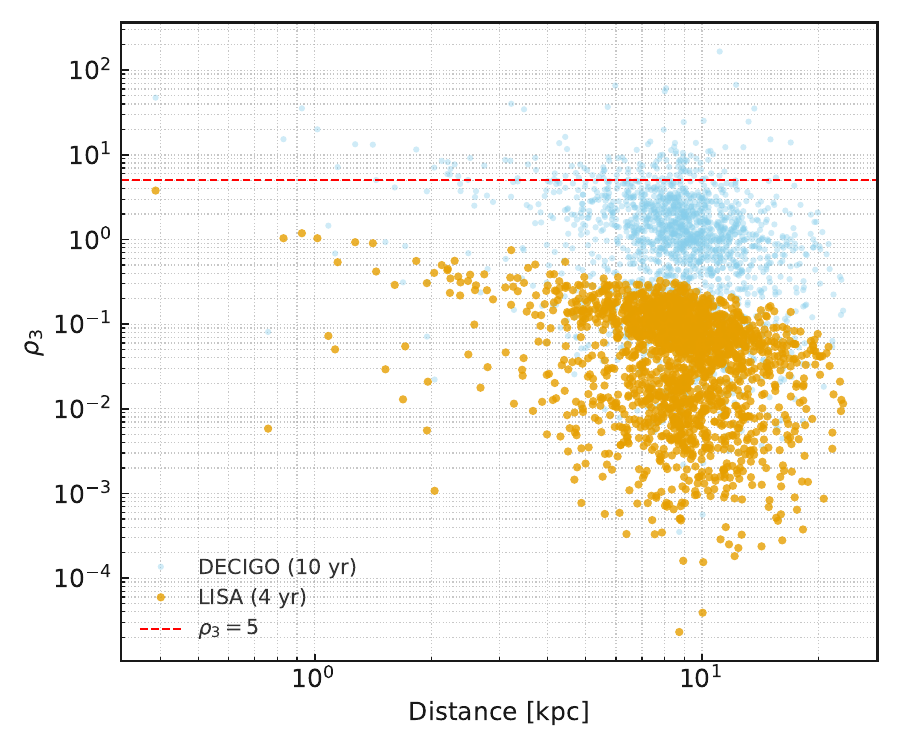}
  \includegraphics[width=0.48\textwidth]{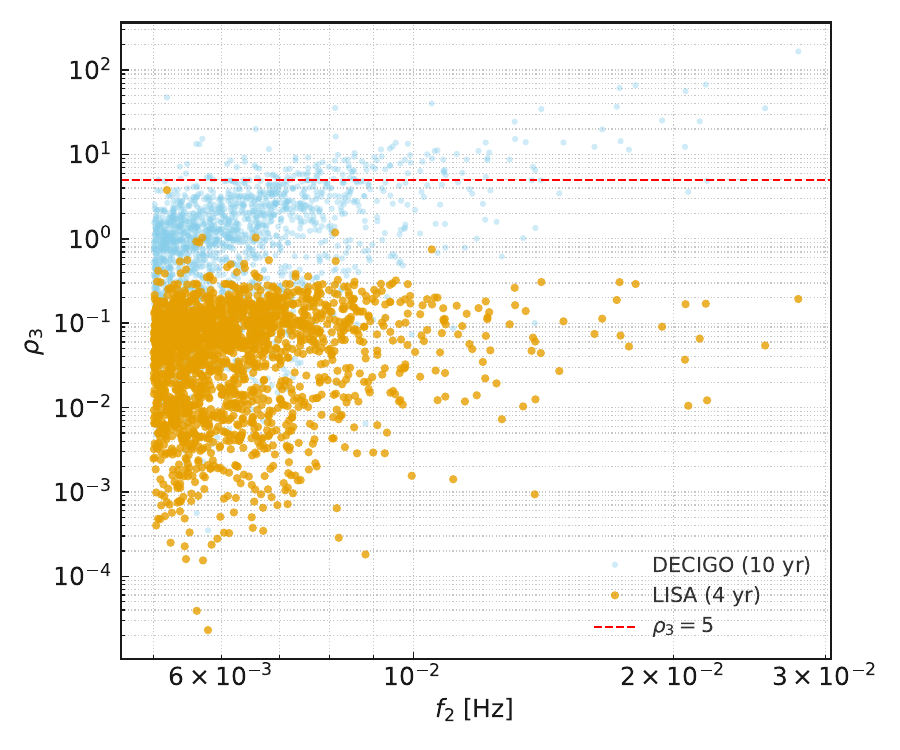}
  \caption{SNRs of the third harmonic ($\rho_3$) as  a function of distance
  (left) and quadrupole frequency $f_2$ (right), for LISA (orange)
  and DECIGO (blue). The horizontal dashed line marks the
  threshold $\rho_3=5$.}
  \label{fig:rho3_scatter}
\end{figure*}

Figure~\ref{fig:rho3_scatter} displays the third-harmonic SNR $\rho_3$ as 
functions of distance (left) and quadrupole frequency $f_2$ (right).
The horizontal dashed line indicates the fiducial threshold ${\rm SNR}=5$.
In the distance panel, the upper envelope of the DECIGO points follows
the expected $d^{-1}$ scaling of the GW amplitude. The detection
fraction decreases gradually with distance: about $56\%$ of systems
within $2.5$~kpc exceed the threshold, and even in the range
$7.5$--$10$~kpc the fraction is still $\sim 6\%$. This demonstrates
that higher harmonics can be probed throughout the Galaxy with DECIGO.
In contrast, the LISA points lie entirely below the threshold line,
confirming that only very nearby binaries would be detectable
with LISA.

In the frequency panel, the DECIGO results exhibit a steep increase of
the third--harmonic SNR $\rho_3$ with the quadrupole frequency $f_2$.
This trend reflects both the post--Newtonian scaling
$\beta \propto f_2^{1/3}$ and, more importantly, the rapid decline of the
extra--galactic foreground $S_{\rm exgal}(f)$ toward higher frequencies,
as shown in Fig.~\ref{fig:noise_exgal}.

{
In our sample population of 1858 binaries at $f_2 > 5\,\mathrm{mHz}$,
this frequency dependence results in 135 systems exceeding the working
threshold $\rho_3=5$, as mentioned earlier.
The detected population is strongly concentrated at the high--frequency
end: 122 of these systems have $f_2 \ge 6\,\mathrm{mHz}$ and 106 have
$f_2 \ge 7\,\mathrm{mHz}$, while 43 systems reside above
$f_2 = 10\,\mathrm{mHz}$.
Although such high--frequency binaries constitute only a small fraction
of the overall population, they contribute a disproportionate share of
the detected systems owing to the rapid growth of $\rho_3$ with
frequency.}

{
This concentration toward higher frequencies indicates that the
main conclusions are not sensitive to the precise choice of the
lower frequency cut adopted in the analysis.
}

The vertical spread of points at fixed $(d,f_2)$ is primarily driven by
the mass asymmetry $\Delta$, consistent with the $\rho_3\propto \Delta$
scaling of higher harmonics, with orientation effects being 
subdominant.

\subsection{Summary}
\label{sec:summary}

In summary LISA will provide a nearly complete census of Galactic
DWDs through detections of the fundamental quadrupole
mode at $f_2\gtrsim5$ mHz. Decihertz observatories such as DECIGO and BBO
will extend this foundation by detecting higher harmonics, in
particular the third mode for hundreds of systems and the first mode
only for rare systems. These capabilities together outline a sequential
observational strategy, progressing from a secure quadrupole census to
detailed multi-harmonic studies in the $\sim 10$ mHz regime.

It should be emphasized that the absolute detection counts depend
to some extent on the assumed population model. Modest changes in the adopted lower-mass distribution lead to variations 
at the level of several tens of systems, without altering the overall
conclusion that the detectable fraction may be on the order of 10\%.
 Further discussion of this dependence is
deferred to Sec.~\ref{sec:limitations}.

We finally note that harmonics induced by orbital eccentricity, as
already noted in Sec.~\ref{sec:signal} and discussed in detail by
\cite{Seto:2025vfg}, lie beyond the scope of this study. At still
higher post-Newtonian order, the amplitudes scale with $\beta^2$, and
even with the very large quadrupole SNRs achievable by DECIGO
($\rho_2^{\rm max}\sim 5.9\times10^4$ in our sample), the corresponding
subdominant contributions are expected to remain undetectable in the
vast majority of systems.

\section{Discussion and Implications}
\label{sec:discussion}

In this section, we discuss the broader implications of our results, including combination with electromagnetic observations.
We also outline the main limitations of our study and directions for
future investigation.

\subsection{Astrophysical implications and EM synergy}
\label{sec:astro}

Electromagnetic information, in particular accurate distances, may be
combined with GW amplitudes $A\propto {\cal M}^{5/3}/d$ to infer the chirp
mass ${\cal M}$, which is less sensitive to finite-size corrections than
chirp-mass estimates based on $\dot f$ \cite{lau2024astrophysical}.
Cross-calibration of these methods would strengthen the robustness of
chirp-mass measurements.

When higher harmonics are detected, the mass ratio can be constrained as
well, enabling estimates of the two component masses. This is
particularly important for AM~CVn systems, where electromagnetic
observations alone often fail to yield solid constraints on mass ratios,
as illustrated by HM~Cnc \cite{kupfer2018lisa,munday2023two}. In such
cases, GW harmonics could provide critical insight into binary configuration and
evolutionary pathways.

\subsection{Limitations and future work}
\label{sec:limitations}

Several limitations of our study should be noted. First, we have
not attempted to model the outspiral phase following Roche-lobe
overflow, focusing instead on the inspiraling regime. As discussed
earlier, the amplitude measurements considered here do not depend directly on
the sign of $\dot f$. However, the quoted fractions (e.g., the $\sim 10$\% for the third harmonic) 
refer specifically to the pre-contact inspiral sample and could differ 
once post-contact outspiral systems are included.

Second, our injection model is based on simplified distributions of
component masses and radii. While this simplified treatment is expected to capture
the main features of the observational prospects, the detailed
quantitative results depend on the adopted population model. For
example, in our baseline configuration ($f_2 \ge 5$ mHz) the working
sample contains 1858 binaries, of which 135 (7.3\%) exceed $\rho_3>5$
in DECIGO. Adopting an alternative but still plausible mass
distribution (He peak shifted to $0.25\,M_\odot$ and the lower bound
raised from $0.11$ to $0.12\,M_\odot$) yields 1897 binaries with 159
detections (8.4\%). These variations illustrate that the absolute
counts should not be over-interpreted. For the assumptions considered
here, the detection fraction is on the order of 10\%.

A further caveat concerns the low-frequency sensitivity of DECIGO. In
the present study, the noise level around 10\,mHz is assumed to be
dominated by radiation-pressure noise in Fabry--Perot arm 
\cite{Kawamura:2011zz}. Because the arm length is relatively short, the
technical requirements on the acceleration noise are demanding, and the
final sensitivity goal may be less ambitious than in the analytic model
adopted here. To assess robustness, we examined the effect of halving
the SNRs. Even under this assumption, the third harmonic remains observable in
$\sim40$ binaries, allowing statistical studies at this level of
degradation. Future design work will clarify realistic targets for the
low-frequency band.

Our LISA forecasts motivate stacking of sub-threshold odd harmonics.
Such methods should be robust to the four-fold phase ambiguity \cite{Seto:2025vud}, guided by the quadrupole posterior, and calibrated with permutation tests.
A systematic study of weighting schemes and noise robustness is left to future work.

\section{Conclusions}
\label{sec:conclusions}

We have investigated the detectability of the first and third
harmonics from Galactic DWDs around 10\,mHz. We conclude
that LISA will provide a nearly complete census of the population
through detections of the quadrupole mode, while decihertz
observatories such as DECIGO and BBO will be able to access higher
harmonics in a significant subset of the same systems. In particular, the third harmonic is expected to be measurable for
roughly 10\% of inspiral binaries with detectable quadrupole emission above 5 mHz.
This will enable constraints on the binary mass ratio.
 Together, these capabilities define a long-term
observational strategy for space-based gravitational-wave astronomy.

While most attention on decihertz observatories has focused on the
0.1--1\,Hz range, where the noise floor is expected to be lowest and
cosmological signals such as an inflationary background may be
probed, our results highlight the importance of the $\sim$10\,mHz
regime, bridging the observational domains of LISA and DECIGO/BBO. {If sensitivities at the level assumed in this work can be realized,
this band will not only connect different mission domains but also
provide valuable astrophysical insights into the configuration and
evolution of compact binaries
\citep[e.g.,][]{PostnovYungelson2014,amaro2023astrophysics}.}

\begin{acknowledgments}
The author thanks the participants of the MIAPbP program
``Enabling Future Gravitational Wave Astrophysics in the Milli-Hertz Regime''
(Munich, 2025) for helpful discussions.
\end{acknowledgments}

\bibliography{ref}
\end{document}